\documentclass[journal,usenames,dvipsnames]{IEEEtran}
\IEEEoverridecommandlockouts

\usepackage{ifpdf}
\ifpdf
   \usepackage{url}
\else
\fi

\usepackage{graphicx}
\usepackage[noadjust]{cite}
\usepackage{hyperref}
\usepackage{caption}
\usepackage{subcaption}
\usepackage{tikz}

\usepackage[usenames,dvipsnames]{pstricks}
\usepackage{epsfig}
\usepackage{pst-grad} 
\usepackage{pst-plot} 

\usepackage{url}
\usepackage[ruled,linesnumbered]{algorithm2e}
\usepackage{flushend}

\begin{document}

\title{Intelligent Anomaly Detection\\ and Mitigation in Data Centers}

\author{\IEEEauthorblockN{Ashkan Aghdai, \, Kang Xi, \, H. Jonathan Chao}
\IEEEauthorblockA{\\NYU Polytechnic School of Engineering}}

\newcommand\copyrighttext{%
  \footnotesize A longer version of this article was submitted to ACM CoNEXT 2014.}
\newcommand\copyrightnotice{%
\begin{tikzpicture}[remember picture,overlay]
\node[anchor=south,yshift=10pt] at (current page.south) {\fbox{\parbox{\dimexpr\textwidth-\fboxsep-\fboxrule\relax}{\copyrighttext}}};
\end{tikzpicture}%
}

\maketitle
\copyrightnotice

\begin{abstract}
Data centers play a key role in today's Internet.
Cloud applications are mainly hosted on multi-tenant warehouse-scale data centers.
Anomalies pose a serious threat to data centers' operations.
If not controlled properly, a simple anomaly can spread throughout the data center, resulting in a cascading failure.
Amazon AWS had been affected by such incidents recently.
Although some solutions are proposed to detect anomalies and prevent cascading failures, they mainly rely on application-specific metrics and case-based diagnosis to detect the anomalies.
Given the variety of applications on a multi-tenant data center, proposed solutions are not capable of detecting anomalies in a timely manner.
In this paper we design an application-agnostic anomaly detection scheme.
More specifically, our design uses a highly distributed data mining scheme over network-level traffic metrics to detect anomalies.
Once anomalies are detected, simple actions are taken to mitigate the damage. 
This ensures that errors are confined and prevents cascading failures before administrators intervene.
\end{abstract}


\section{Introduction}\label{sec_intro}
Data centers are the most critical infrastructure to host large-scale, complex enterprise and cloud applications \cite{barroso2009datacenter}.
Over the years, data centers have undergone great growth in two aspects.
On one hand, the hardware capacity of a data center has become much more powerful than ever before.
On the other hand, the popularity of cloud computing has motivated the growth of multi-tenant support in data centers \cite{wang2008study, shen2011cloudscale, miller2008microsoft}.
As a result, it is typical that today's data centers host tens or even hundreds of tenants and that each tenant runs tens or hundreds of applications.
With such unprecedented system scale and application complexity, there comes a great challenge: how to deal with anomalies caused by a variety of factors, such as hardware faults, configuration mistakes, software bugs, flaws in application designs, and attacks.

This paper addresses large-scale anomaly detection and mitigation. In such cases an anomaly in a particular device or application could ``spill off'' to affect many other applications in the same data center.
In the worst case, such an anomaly could trigger a sequence of problems and lead to system-wide outages, known as a cascading failure \cite{dobson2007complex, nedic2006criticality, brewer2001lessons}.
For multi-tenant data centers, a cascading failure could affect a large number of tenants and result in considerable loss.
The Amazon AWS outage in April 2011 is a typical example of cascading failure \cite{a1}.
During a routine maintenance, a routing misconfiguration created congestion on a small number of links.
The congestion caused loss of keep-alive messages between many primary and backup storage nodes.
As a results, the primary nodes mistakenly decided that the backup nodes were down, and thus started to create new mirror nodes.
The mirroring operation intensified the congestion and subsequently triggered even more mirroring operations, which in turn exhausted the data center storage and bandwidth resources.
Furthermore, the above events triggered such a large number of messages to the control plane that the load eventually brought down the control plane.
The outage of the control plane caused even more impacts to the entire system.
Although the initial misconfiguration was detected and fixed quickly, the damage had been made and the outage affected a large number of tenants running on top of AWS.
It took four days to completely restore the entire system.

Due to the complexity of large-scale data centers and applications, it is often non-trivial to detect an anomaly, locate the root causes, and perform mitigation.
While insightful discoveries have been made on data center traffic patterns that suggest regardless of applications running in a data center some trends tend to be consistent \cite{kandula2009nature, benson2010network, aghdai2013traffic}, current solutions heavily rely on case-specific analysis and are often labor-intensive \cite{wu2012netpilot, yu2011profiling}.
One must have certain knowledge on the type and characteristics of an application to monitor its operation and perform anomaly detection.
While such approaches are indispensable, it is not the best choice to solely rely on experienced system administrators to perform the tasks, especially in large-scale systems such as a data center.
In particular, it is often impossible for the operator of a multi-tenant data center to obtain the type and characteristics of a tenant's applications.

According to a recent analysis \cite{cloudfail}, from 2009 to 2011 the number of cloud vulnerability incidents more than doubled.
In particular, some disruptions were so severe that they caused large-scale service outage lasting for hours or even days.
For example, the Amazon AWS experienced several large-scale outages \cite{armbrust2010view}.
Reddit and NetFlix are just a few example of web services that experienced wide disruptions during recent Amazon AWS incidents \cite{a2, a3}.
Other major cloud players such as Microsoft and Google have experienced similar incidents as well \cite{mi1, g1}. 
Table \ref{Outages} lists a few well-documented major data center failures in the past few years.
This list is by no means exhaustive.
Many service providers even choose not to report incidents for various reasons.
We estimate that the actual impacts from data center anomalies are far more intense than what has been reported.

We propose to design a highly resilient data center with the following key properties:

\begin{itemize}
\item \textbf{Detect anomalies as quickly as possible.}
Errors should not propagate through the entire data center.
Detecting anomalies in real time enables data center operators to proceed with quick mitigation actions.
This will prevent the cascading failures so as to confine the damage that could possibly caused by a small application or network error.

\item \textbf{Avoid using application-specific information for detection.}
Multi tenancy, widely adapted in warehouse-scale data centers, limits amount of information that is available to data center operators.
For example, detecting a misconfiguration in a local name resolution service is pretty straightforward by analyzing its logs.
However, these logs are only available to the tenant.
Thus, data center operators cannot rely on application-specific metrics for anomaly detection.
A solid and useful anomaly detection scheme should make the most from information accessible to data center operators.

\item \textbf{Mitigate errors using simple actions.}
Our analysis of previous failures shows that examples of cascading failure errors could easily be prevented by using simple mitigation actions such as throttling flows in the data center.
Such simple actions are effective to prevent errors from spreading to the entire system and do not generate severe negative impact to the system.
\end{itemize}

\begin{table}[t]
\centering
\begin{tabular} {c c c c}
\hline
Service & Date & Duration & Refer to\\
\hline
Amazon AWS & Apr. 2011  & 4 Days & \cite{a1}\\
Amazon AWS & Oct. 2012 & 8 Hours & \cite{a2}\\
Amazon AWS & Dec. 2012 & 12 Hours & \cite{a3}\\
Sidekick & Fall 2009 & 1 Week & \cite{s1}\\
Hotmail & Dec. 2010 & 3 Days & \cite{mi1}\\
Intuit & Jun. 2010 & 3 Days & \cite{i1}\\
Google services & Jan. 2014 & 30-60 Minutes & \cite{g1}\\
\hline
\end{tabular}
\caption{Recent well-documented data center outages}
\label{Outages}
\end{table}

\subsection*{Contributions}

We propose an automated anomaly detection and mitigation scheme for data centers.
Our scheme uses data mining over network-level traffic metrics to detect anomalies.
We also study several cases of well-documented outages in Amazon AWS.

A multi-stage detection and mitigation scheme is used to prevent propagation of errors in data centers.
During the first stage, our algorithm detects traces of abnormal behavior in specific traffic metrics.
At the next stage, the correlation between detected traces of anomalies will be determined.
Those with high correlation are potentially anomalies.
Once an abnormality is detected, it is essential to perform mitigation actions so as to avoid potential cascading failures.
Small-scale and local anomalies are neglected as long as they do not spread.

\subsection*{Organization}

The rest of the paper is organized as follows.
\hyperref[Sec_Frm]{\S \ref*{Sec_Frm}} introduces our design for anomaly detection and mitigation in data centers.
\hyperref[sec_sym]{\S \ref*{sec_sym}} defines and explores symptoms of abnormal behavior in traffic patterns.
\hyperref[sec_2s]{\S \ref*{sec_2s}} elaborates the proposed 2-step anomaly detection algorithm.
\hyperref[DC_Analysis]{\S \ref*{DC_Analysis}} explores well-documented outages in Amazon AWS.
\hyperref[sec_sim]{\S \ref*{sec_sim}} explores existing work in this area.
Finally, \hyperref[sec_conc]{\S \ref*{sec_conc}} concludes the paper.


\section{Anomaly Detection and \\ Mitigation Framework} \label{Sec_Frm}

In this paper, we target high-impact anomalies that pose threat to the data center at large scale.
That is, the anomaly affects many VMs or services.
Traffic traces from affected VMs are influenced during the anomaly.
We show that certain traffic metrics are influenced the most during anomalies.

To detect anomalies, our design monitors traffic traces.
VMs with suspicious traffic patterns are further analyzed.
Our algorithms are designed to find groups of VMs that suffer from the same anomaly.
Upon detection of such groups, data center operators are immediately notified.
Furthermore, mitigation actions are performed so as to prevent anomalies from spreading through the data center.

\subsection{Symptoms}

Our scheme constantly monitors traffic traces looking for \emph{symptoms} of anomalous behavior.
A symptom is a certain type of pattern in traffic traces, such as an increase in Round Trip Time (RTT), or abrupt change in creation/termination rate of flows.
We base our detection on observing and analyzing symptoms.
A large-scale anomaly that affects a lot of VMs in a data center creates a good deal of symptoms.
In particular, the same type of symptoms tend to be observed in large quantity at many hosts.
For example, a misconfigured router may cause increased RTT and packet losses for many flows.
A malfunctioning cloud application may cause many of its VMs to create new flows at an abnormal speed.
Generated symptoms caused by a specific anomaly are \emph{correlated} in time domain as well as space domain.
More specifically, an anomaly creates symptoms with the same pattern over time and the symptoms appear on malfunctioning hosts at roughly the same time.
These two types of relations between the observed symptoms are called time domain and space domain \emph{correlation} respectively.

Our design relies on global scale correlation-based analysis of the symptoms to detect anomalies.
While it is true that certain anomalies cause application outages but do not generate obvious symptoms, neither do such anomalies affect the data center operation at large.
Detection and mitigation of such anomalies are beyond the scope of this work.

\subsection{Architecture}

\begin{figure}[t]
\centering
\includegraphics[keepaspectratio=true,width=0.45\textwidth]{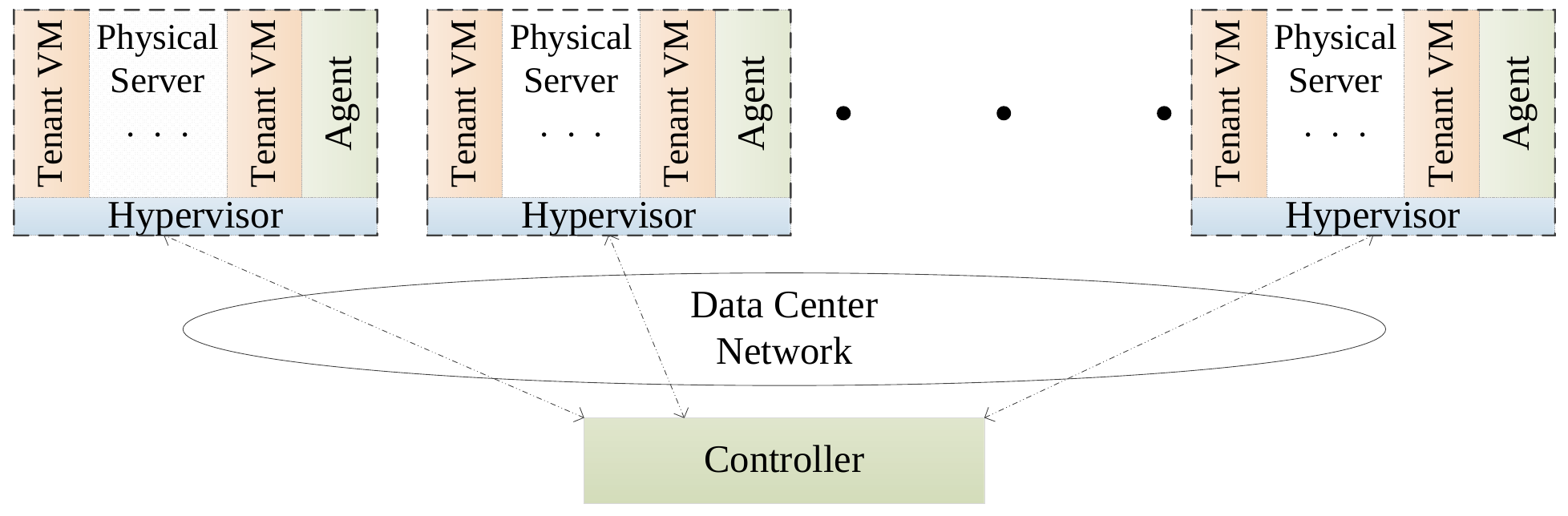}
\caption{Proposed architecture for anomaly detection.}
\label{Arch_fig}
\end{figure}

Figure \ref{Arch_fig} illustrates the architecture of our design.
The main components are Agents and the Controller.

\begin{itemize}
\item \textbf{Agent}:
An agent is a VM that runs on a physical server hosting application VMs.
The agent monitors and analyzes the inbound and outbound traffic to identify various types of symptoms.
Agents communicate with controller to report detected symptoms.
Hypervisors are modified to recognize the special role of the agent.
Thus, they interact with agents accordingly.
At fine granularity, the hypervisor duplicates the header of each inbound and outbound packet and passes the header to the agent for analysis.
Packet duplication at hypervisor ensures that our scheme does not incur any additional delay in packet forwarding.

\item \textbf{Controller}: Controller receives symptoms' reports from the agents and performs the correlation.
It make decisions on anomaly detection based on the global correlation of symptoms in time and space domains.
Controller is responsible for raising alerts to data center operators and having hypervisors initiate the automated mitigation.
Mitigation is done by limiting the resource access (typically bandwidth) of certain type of traffic or certain VMs.
\end{itemize}

The design can scale out by using multiple controllers.

\subsection{Detection Method}

A specific anomaly triggers similar symptoms in affected hosts.
These symptoms are observed at affected hosts at roughly the same time.
Therefore, they are correlated in space domain.
Similarly hosts affected by the same anomaly experience the same pattern of symptoms.
Hence, symptoms are also correlated in time domain.

For example, a link congestion causes increased round trip times followed by increased number of packet drops.
As a result, affected hosts eventually experience large number of packet drops.
Spatial correlation can detect these hosts.
During the anomaly, all of the affected hosts experience increased RTT as well.
Therefore, time-series correlation can analyze spatial correlation results and find group of hosts that show correlation in time domain as well as space domain.

In our framework, agents periodically monitor and analyze traffic traces and summarize observed symptoms into a report.
Controller correlates reports in space domain to reveal clusters of hosts that experience similar symptoms at roughly the same time.
Then, the controller performs time domain analysis on each of the spatially-correlated clusters.
Groups of hosts with high spatial as well as time domain correlation are considered to be abnormal.
These groups are reported to the data center operators and are marked for performing mitigation.

\section{Symptoms} \label{sec_sym}

\begin{table*}[t]
\centering
\begin{tabular} {c c c c c}
\hline
RTT & RST & Duplicate Acks & Active Flows & Flows Size \\
\hline
Similarity & Frequency & Frequency & Fluctuations & Fluctuations \\
\hline
\end{tabular}
\caption{Methods to generate elemnts in SR.}
\label{report_generation}
\end{table*}

We adopted the following set of symptoms for anomaly detection:

\begin{itemize}

\item \emph{Change of active flows:}
Sudden change in creation/termination rate of flows is a symptom of abnormal behavior.
The key observation is that the creation of new flows by a large number of hosts in a short duration is often a suspicious phenomenon that needs careful monitoring and analysis.

\item \emph{Change of flow size:}
Abrupt change in average flow size is another symptom of suspicious application activities.
Due to the nature of general applications, the flow size (the total amount of data exchanged) often has consistent patterns in both the time domain and spatial domain.
A sudden change of the attribute either indicates a radical change of the application or an anomaly.

\item \emph{Similarity of observed RTTs:}
A sudden change in pattern of observed round trip times between a pair hosts compared to their history, is another symptom of potential anomalies. 
The RTT between a pair of hosts is determined by several factors: propagation delay, processing delay in switches and routers, queuing delay in switches and routers, and processing delay at the receiver.
In data centers, the main factors to consider include the queuing delay and the processing delay.
A dramatic increase of the queuing delay is mostly caused by congestion, while an increase of the processing delay can be caused by overloading or malfunctioning at the receiver.
While a small number of isolated RTT increases do not pose any threat, a large number of concurrent such symptoms could indicate a critical problem that requires attention.

\item \emph{Number of duplicate TCP ACKs:}
Observing large number of triple duplicate acknowledgments in a TCP flow is another symptom of potential anomaly in networks.
Duplicate TCP ACKs are generated if some TCP segments are dropped in the network.
This symptom can be observed during normal operation because occasional packet dropping is normal and could even be caused by certain algorithms such as random early detection (RED) \cite{floyd1993random}.
However, a network problem such as prolonged congestion or misconfiguration could trigger an unusual number of such symptoms in large scale.

\item \emph{Number of TCP RST signals:}
Observing TCP RST signals is another symptom indicating anomalous behavior in applications or operating system's protocol stack. 
TCP reset (RST) is a one-bit flag in the TCP header \cite{postel1981transmission}.
A host sends out a TCP RST if it notices that it is not receiving acknowledgements for anything it sends and thus the connection is unsynchronized.
The signal could also be triggered by certain middleboxes.
In normal operation, such flags should be scarce. Therefore, this symptom is a good indicator of abnormal behavior.

\end{itemize}

\subsection{Generation of Symptom Reports}

Each agent reports to the controller one or multiple metrics of an observed symptom. The metrics are:

\begin{itemize}

\item \emph{Frequency:}
Agents count the number of occurrences for each symptom.
This metric is used for observed triple duplicate acknowledgments and RST flags.
These incidents by themselves indicate an abnormal behavior in TCP connections, therefore no further processing is needed for these types of symptom.

\item \emph{Fluctuations:} 
Local agents observe average size of a symptom in each time window.
The signed relative change of the attribute in comparison to that of the previous time window is calculated and reported in SR:
$\Delta X = \frac{X_{Current Sample} - X_{Previous Sample}}{X_{Current Sample}}$

This metric is used for detecting change in number of active flows and change in average size of flows.

\item \emph{Similarity:}
Local agents keep track of observed symptoms in each time window.
The objective is to determine how similar the current set of observed symptoms is to that of the previous time window.
The calculated similarity is reported in SR.

\begin{figure}[b]
\centering
\includegraphics[keepaspectratio=true,width=0.45\textwidth]{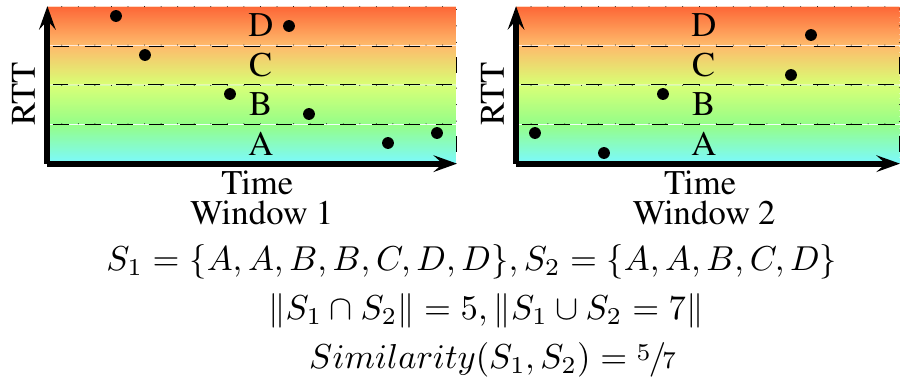}
\caption{An example of calculating similarity.}
\label{RTT_Sample}
\end{figure}

We use Jaccars's similarity coefficient \cite{rajaraman2012mining} to calculate the similarity between observed RTTs.
Jaccard's similarity coefficient measures similarity between finite sample sets, and is defined as the size of the intersection divided by the size of the union of the sample sets.
To calculate the Jaccardian similarity of current time window to that of the previous window, agents dequantize observed symptoms into labels.
Jaccardian similarity is calculated between current set of labels and the set from the previous time window.
This value is reported as the similarity of observed symptoms to those of the previous time window.

This metric is used for detecting changes in RTT.
Due to the nature of physical communication at network observed observed RTTs do fluctuate for a flow.
However, a simultaneous increase in observed RTTs may indicate a problem with the network and/or applications.
Figure \ref{RTT_Sample} illustrates an example of this process.
In this example observed RTTs from two consecutive time windows are shown using dots.
Numbers are converted to labels using four bands.
Finally Jaccardian similarity between the two sets is calculated and reported.

\end{itemize}

Table \ref{report_generation} presents the method used for each symptom.

\section{Symptom Analysis for Anomaly Detection} \label{sec_2s}

Our detection scheme relies on two algorithms to reveal correlation between observed symptoms in both space and time domains.

\subsection{Distributed Analysis of SRs\\ in Space Domain (DAS)}

Spatial correlation is performed using SR vectors.
VMs with similar SRs are considered to have correlation.
From this point of view, the problem of finding spatial correlation is quite similar to clustering problem in statistical data analysis \cite{rajaraman2012mining, hartigan1979algorithm, steinbach2000comparison}.
That is, given a set of vectors, identify one or more groups in such a way that vectors in the same group are more similar to each other than to those not belonging to that group.
The clustering problem has been extensively studied in statistical data analysis and has been applied in a variety of fields such as data mining \cite{rajaraman2012mining} and machine learning \cite{rajaraman2012mining}.

DAS is a distributed modification of K-Means clustering \cite{hartigan1979algorithm}.
DAS has 2 components, the first component is performed at each agent.
It outputs local VMs that show similar symptoms.
Second component, running at controller, has a global view of VMs.
It receives SRs that shown similar symptoms, and outputs VMs with similar symptoms at data center scale.

\begin{algorithm}[b]
  \SetAlgoLined\DontPrintSemicolon
  \SetKwFunction{algol}{DAS at agent}
  \SetKwProg{myalgl}{Algorithm}{}{}
  \myalgl{\algol{SR}}{
  $P_{0} \leftarrow \bar{0} $\;
  $P_{i:1..5} \leftarrow i$th. unity vector \;
  $C_{i:0..5} \leftarrow \emptyset $\;
  \While{There are SRs left}{
    Read $SR_j$\;
    $D_{i:0..5} \leftarrow$ Distance($SR_j$, $P_i$)\;
    $k \leftarrow$ ArgMin($D_{i:0..5}$)\;
    $C_k.$Add($SR_j$)\;
   }
   \Return{$C_{i:1..5}$}\;
  }
  \caption{DAS at Agents}  \label{DAS_agent}
\end{algorithm}

\subsubsection{DAS at Local Agents}

Agents send VMs with similar SRs to controller, filtering out normal VMs. 
In order to perform this task, agents run modified clustering with $K = 6$ clusters.
Algorithm \ref{DAS_agent} performs clustering with predefined cluster representatives.
It finds closest cluster representative, $P_{i:0..5}$ to each SR and puts the SR in appropriate cluster, $C_{i:0..5}$.
Euclidean distance \cite{rajaraman2012mining} is used as the metric.
$P_0 = \bar{0}$, indicates experiencing no symptom.
$P_{i:1..5}$ is the $i$th. unity vector in $R^5$, which indicates experiencing $i$th. symptom.
VMs showing no specific symptom will be clustered in default cluster, with representative: $P_0$.
While, potentially abnormal VMs are clustered according to their dominant symptom.
For VMs clustered in $C_{i:1..5}$ their SRs are sent to controller for further analysis.

\subsubsection{DAS at Controller}

Main part of clustering is done at the controller.
Controller periodically receives all SRs showing symptoms.
It runs a highly modified K-Means clustering, Algorithm \ref{DAS_ctrlr}, to find clusters of VMs that experience similar symptoms simultaneously.
At each iteration of clustering, SRs are clustered into $K_i$ clusters.
It starts with $K_0 = 6 * L$ where L is number agents.
At each iteration, number of clusters is halved: $K_i = K_{i-1}/2$.
To find optimal number of clusters, we use an elbow criterion \cite{zu2002analysis}.
At each iteration with $K = K_i$, clustering cost, $CC_i$ is defined to be $CC_i = \frac{CG_i}{K_i}$.
Where $CG_i$ indicates clustering gain, defined as $CG_i = \bar{CS}_{i} - \bar{CS}{i-1}$ where $\bar{CS}_{i}$ denotes average cluster size at $i$th. iteration. 
Clustering is continued while $CC_i$ is improved.
It is stopped when $CC_i$ is in the 5\% range of $CC_{i-1}$, or $K_i$ reaches one.

\begin{algorithm}[t]
  \SetAlgoLined\DontPrintSemicolon
  \SetKwFunction{algo}{DAS at Controller}
  \SetKwProg{myalg}{Algorithm}{}{}
  \myalg{\algo{SR}}{
   $K \leftarrow 6L$ \;
   $i, CS_0, CC_0, \Delta CC \leftarrow 0$ \;
   \While{$K > 0$ and $\Delta CC < 0.95$}{
    Clusters$_{K} \leftarrow$ Cluster(SR, $K$) \;
    $i \leftarrow i + 1$ \;
    $CS_i \leftarrow$ Clusters$_{K_i}.$AverageSize \;
    $CC_i \leftarrow \frac{CS_i - CS_{i-1}}{K_i}$\;
    $\Delta CC \leftarrow \frac{CC_i - CC_{i-1}}{CC_{i-1}}$ \;
    $K \leftarrow$ Floor($K/2$) \;
   }
   \Return{Clusters$_K$}\;
  }
  
  \setcounter{AlgoLine}{0}
  
  \SetKwFunction{proc}{Cluster}
  \SetKwProg{myproc}{Function}{}{}
  \myproc{\proc{SR, K}}{
   $P_{i:1..K} \leftarrow$ K Random SRs \;
   Clusters$_{NEW} \leftarrow \emptyset$ \;
   \Repeat{Clusters$_{NEW} =$ Clusters$_{OLD}$}{
    Clusters$_{OLD} \leftarrow$ Clusters$_{NEW}$ \;
    Clusters$_{NEW}, C_{i:0..K} \leftarrow \emptyset$ \;
    \While{There are SRs left}{
     Read $SR_j$\;
     $D_{i:0..K} \leftarrow$ Distance($SR_j$, $P_i$)\;
     $k \leftarrow$ ArgMin($D_{i:0..K}$)\;
     $C_k.$Add($SR_j$)\;
    }
    Clusters$_{NEW}.$Add($C_{i:1..K}$) \;
    $P_{i:1..K} \leftarrow$ $C_i.$Mean \;
   }
   \Return{Cluster$_{NEW}$}\;
  }
  
  \caption{DAS at Controller}  \label{DAS_ctrlr}
\end{algorithm}

\begin{table*}[]
\centering
\begin{tabular} {c c c c c c}
\hline
\textbf{Anomaly Type} & \textbf{RTT} & \textbf{RST} & \textbf{Duplicate Ack}s & \textbf{Flow Size} & \textbf{Active Flows} \\
\hline
Router misconfig (AWS, Apr. 12) \cite{a1} & increase & increase & increase & large & increase\\
\hline
DNS misconfig (AWS, Oct. 12) \cite{a2} & increase & & & small & increase\\
\hline
Load balancer misconfig (AWS, Dec. 12) \cite{a3} & increase & increase & increase & & change \\
\hline
\end{tabular}
\caption{Analysis of recent well-documanted incidents at Amazon AWS}
\label{case_study}
\end{table*}

\subsection{Time Domain Analysis of Clusters (TAC)}

Controller further scrutinizes the historical records of spatially-related VMs to perform time domain correlation.
There are two objectives to conduct this operation.
Firstly, it is possible that a group of hosts happen to exhibit similar symptoms even though the symptoms are simply random and isolated events.
Although DAS filters out many coincidences, it is not possible to eliminate such false alarms.
Therefore, we need to perform time domain correlation to look back to the history to see whether the symptoms are persistent or ephemeral.
Secondly, anomalies tend to trigger multiple symptoms in a sequential way.
For example, overloading on a link leads to increased RTT for many flows, and persistent overloading further triggers duplicate ACKs and possibly TCP RST signals.

The time domain correlation is performed using Time Domain Analysis of Clusters (TAC).
TAC is performed on the output of DAS.
Initially, agents send 5 most recent reports of all the hosts in DAS output clusters to the controller.
Supposing symptom $S_j$ has been detected among a set of hosts using space correlation, we set a look-back window and observe the occurrence of all the symptoms $S_1$, $S_2$, ..., $S_k$ (not including the final detected symptom by DAS) in the window for each host.
The result for host $i$ in the cluster, $h_i$, is a vector $\Gamma_i = [\alpha_{i, 1} , \alpha_{i, 2} , ..., \alpha_{i, k}]$, where $\alpha_{i, k}$ is the indicator for symptom $S_k$ at host $h_i$.
Then, pairwise Jaccardian similarity is calculated between all $\Gamma_i$ vectors.
If vectors $\Gamma_i$ and $\Gamma_j$ have similar values for $\alpha_{i, k}$ and $\alpha_{j, k}$, it means that symptom $S_j$ has been persistent and.
In addition, hosts i and j in the set have exhibited at least one symptom $S_k$ in the first place since they have been outputted by DAS.

TAC outputs VMs with high similarity in their pair-wise $\Gamma$-vector.
These VMs experienced at least one similar symptom at roughly the same time while showing similar pattern of symptoms in their provided history of time windows.
Therefore, they have shown correlation both in space domain and time domain and are abnormal.

\subsection{Mitigation}

The mitigation process includes two steps.
The first step is to classify the flows or VMs to different groups based on their symptoms.
The goal is to distinguish misbehaving flows and VMs that worsen the problem from victim flows and VMs that are suffering from the misbehaving ones.
For example, while many flows may be suffering from increased RTTs, a specific type of flows are reported to be increasing in quantities.
So it is very likely that this group of specific flows are taking much of the bandwidth and overloading the networks, causing other flows to suffer from congestion.
The second step is to limit bandwidth allocation to the misbehaving flows and VMs. 
Therefore, we will not simply block the traffic of all the suspicious flows and VMs.
Instead, we will selectively throttle the bandwidth of VMs that shown abnormal behavior.


\section{Case Study} \label{DC_Analysis}

We have further elaborate our design by analyzing how well it would have worked on well-explored large scale data center outages.
Table \ref{case_study} lists cases of recent reported large-scale anomalies at Amazon AWS.

The first case happened in April 2011 \cite{a1}, where a mistake in routing caused data center wide congestions.
This incident was explored in detail in \S1.
During the incident, primary nodes started to create new mirror nodes, which affects average flow sizes shifting it towards larger flows.
At this stage we expect agents to create large number of reports triggered with change in flow size.
This correlation in space domain will create clusters of hosts experiencing this phenomena.
As anomaly continues link congestions start to appear.
Appearing link congestion increases the RTT as well as number of triple duplicate acks.
At this stage agents report large number of hosts that experience increase in RTT with large number of triple duplicate acks.
Controller receives the second wave of reports that belong to a portion of hosts that experienced space domain correlation.
Therefore, clusters experience similar patterns of symptoms and show time domain correlation as well.
Hence, as link utilizations increase our scheme will discover space domain as well as time domain correlation.
Automated mitigation will throttle flows from affected hosts which in turn alleviates congestion on links.

Our design is able to detect this anomaly and keep the network alive by throttling flows.
By notifying system administrators in a timely manner, it will provide sufficient time for them to find the root cause.
Meanwhile basic services are kept alive thanks to the automated mitigation actions.

In the second case, happened at October 2012, a bug in an operational data collection agent degraded Amazon Elastic Compute Cloud (EC2) and Amazon Elastic Block Store (EBS) performance.
In some cases, EBS volumes became stuck \cite{a2}.
The root cause for this problem goes back to 1 week earlier when one of the data collection servers was replaced.
While not noticed at the time, the DNS update did not successfully propagate to all of the internal DNS servers, and as a result, a fraction of the storage servers did not get the updated server address and continued to attempt to contact the failed data collection server \cite{a2}.
Because of the design of the data collection service (which is tolerant to missing data), this did not cause any immediate issues or set off any alarms.
However, the inability to contact a data collection server triggered a latent memory leak bug in the reporting agent on the storage servers.
Rather than gracefully dealing with the failed connection, the reporting agent continued trying to contact the collection server in a way that slowly consumed system memory until it reached a critical mass and caused some of the provisioning services to become stuck eventually.
During the period where storage servers tried to reach the missing server, we expect a lot of flows to be created and terminated rapidly.
This, translates to increase in flow creation rate, huge number of triple duplicate acks, and shift in average flow size to smaller flows.
Therefore, this complex anomaly caused by a misconfiguration in DNS entries, could have been detected after replacing the data collection server.
Although our scheme is not able to pinpoint to the memory allocation bug, it is able to detect the original anomaly in a timely manner.

In another example, during a maintenance in December 2012 \cite{a3} human error caused disruptions and down times in services that use Amazon Elastic Load Balancing.
On a group of load balancers, state data was partially deleted by a maintenance process.
As this continued, some customers began to experience performance issues with their running load balancers.
During the anomaly, partial loss in load balancer's state data caused severe changes in traffic matrix.
Therefore, VMs may have experienced overload or underload which directly affects number of active connections.
Overloaded VMs cause increased average processing time for replying messages.
As a increase in RTT would have been observed, had our scheme were used at Amazon.
Also it is safe to assume that overloaded VMs started to drop packets resulting in increased number of triple duplicate acks and TCP RSTs.
Therefore based on the huge number of correlated symptoms that were generated, this anomaly would have been detected in a matter of tens of seconds using our design.

\section{Related Work} \label{sec_sim}

Anomaly detection and mitigation in data centers have recently received a lot of attention \cite{yu2011profiling, wu2012netpilot, tan2012prepare, arefindiagnosing, sun2011identifying, zhou2011tap, attariyan2012x, jeswani2012adaptive}.
However, research in this area is still in its early stage.
The most relevant works to our design include include SNAP \cite{yu2011profiling}, NetPilot \cite{wu2012netpilot}, PREPARE \cite{tan2012prepare}, and FlowDiff \cite{arefindiagnosing}.

SNAP (Simplified Network-Application Profiler) is designed to perform data analysis offline for identifying misbehaving applications.
It collects TCP statistics and socket-call logs from all the hosts and performs centralized correlation to identify application flaws.
However, typical data center operators may not have access to individual hosts and thus could not get the TCP statistics and socket-call log files.
This is often the case for multi-tenant data centers.
SNAP does not address real time detection and mitigation of anomalies.

NetPilot is designed to automate data center network failure mitigation.
Its objective is to quickly mitigate failures rather than finding a root cause and ultimately resolving the problem.
The mitigation process is the same as those adopted by today's network operators: deactivating or restarting offending components.
However, NetPilot is unique in that it takes an intelligent trial-and-error approach to automate the process, including an Impact Estimator that helps guard against overly disruptive mitigation actions and a failure-specific mitigation planner that minimizes the number of trials.
However, NetPilot is unable to deal with complex errors caused by application/network/devices misconfigurations. 

PREPARE is designed to provide anomaly prevention for virtualized cloud computing infrastructures.
In other words, preventive measures are taken to prevent applications from experiencing any anomaly. 
The proposed anomaly prediction scheme relies on system-level metrics such as CPU, memory, and disk I/O usage, and adopts a 2-dependent Markov model in conjunction with a tree-augmented Bayesian networks (TAN) model that is used as the classifier Once an alarm detected, PREPARE identifies faulty VMs and steers the system away from the violation state.
The scheme also performs elastic VM resource scaling and VM migration.
PREPARE does not analyze traffic characteristics, as it is designed to detect application errors.
It does not address the errors caused by network device failures or network misconfigurations.

FlowDiff collects information from all entities involved in the operation of a data center and builds behavioral models to detect operational problems in the data centers.
By comparing the current model with pre-computed and known-to-be-stable models, FlowDiff detects operational problems ranging from host and network failures to unauthorized accesses.
It relies on passive measurements on control traffic from programmable switches to a centralized controller in a software-defined networking (SDN) environment.
The design consists of two phases: modeling and diagnosis.
In modeling, FlowDiff creates signatures for applications using flow attributes, connectivity graph and delay distribution of packets belonging to the flows from specific application such as domain name systems (DNS).
During diagnosis, FlowDiff looks for changes in application and/or infrastructure signatures and raises alerts.
However, FlowDiff does not perform automatic mitigation.
Instead, its purpose is to alert the operator.
Also FlowDiff targets only data centers that are based on SDN.


\section{Conclusions} \label{sec_conc}

We propose a novel application-agnostic scheme to rapidly detect and mitigate anomalies in data centers.
We use distributed processing and data mining techniques to detect anomalies in a timely manner.
Detected anomalies are mitigated using simple actions such as throttling flows that belong to abnormal VMs.
Mitigation ensures that anomalies do not spread throughout the data center.
Our solution not only prevents cascading failures, but also provides sufficient time and information for data center operators to locate the root cause of the anomaly and resolve it with a profound solution.

\small{\bibliographystyle{abbrv}
\bibliography{Reference}}

\end{document}